\documentclass[
reprint,
superscriptaddress,
showpacs,
nofootinbib,
 amsmath,amssymb,
 aps,
 prd,
 eqsecnum,
]{revtex4-1}

\usepackage{graphicx}
\usepackage{color} 
\usepackage{dcolumn}
\usepackage{bm}

\newcommand{\refeq}[1]{Eq.~\eqref{#1}}
\newcommand{\refeqs}[2]{\textrm{Eqs}.~(\ref{#1})-(\ref{#2})}

\newcommand{\re}{\text{Re}}
\newcommand{\tr}{\text{Tr}}
\newcommand{\lwrsim}{\raise0.3ex\hbox{$<$\kern-0.75em\raise-1.1ex\hbox{$\sim$}}}

\begin{document}


\title{Discretization effects on renormalized gauge-field Green's functions, \\ scale setting and gluon mass}

\author{Ph.~Boucaud} 
\affiliation{ Laboratoire de Physique Th\'eorique (UMR8627), CNRS, Univ. Paris-Sud, Universit\'e Paris-Saclay, 
91405 Orsay, France}
\author{F.~De Soto}
\affiliation{Dpto. Sistemas F\'isicos, Qu\'imicos y Naturales, 
Univ. Pablo de Olavide, 41013 Sevilla, Spain}
\author{K.~Raya} 
\affiliation{Department of Integrated Sciences and Center for Advanced Studies in Physics, Mathematics and Computation; 
University of Huelva, E-21071 Huelva; Spain.}
\affiliation{School of Physics; Nankai University, Tianjin 300071, China}
\author{J.~Rodr\'{\i}guez-Quintero}
\affiliation{Department of Integrated Sciences and Center for Advanced Studies in Physics, Mathematics and Computation; 
University of Huelva, E-21071 Huelva; Spain.}
\affiliation{CAFPE, Universidad de Granada, E-18071 Granada, Spain}
\author{S.~Zafeiropoulos}
\affiliation{Institute for Theoretical Physics, Heidelberg University, Philosophenweg 12, 69120 Heidelberg, Germany}




\date{\today}

\begin{abstract}

The $SU(3)$ gauge-field propagators computed from the lattice have been exhaustively used in the investigation of the low-momentum dynamics of QCD, in a judicious interplay with results from other nonperturbative approaches, and for the extraction of fundamental parameters of QCD like $\Lambda_{\overline{\rm MS}}$ as well. The impact of the discretization artifacts and their role in the extrapolation of the results to the continuum limit have not been fully understood so far. We report here about a very careful analysis of the physical scaling violation of the Landau-gauge propagators renormalized in MOM scheme and the Taylor coupling, steering us towards an insightful understanding of the effects from discretization artifacts which makes therefore possible a reliable continuum-limit extrapolation.

\end{abstract}

\pacs{12.38.Aw, 12.38.Lg} 
\maketitle


\section{\label{sec:intro} Introduction}

The understanding of the infrared (IR) dynamics of Quantum Chromodynamics (QCD) has been very much boosted in the last decade, mainly owing to the consistent convergence of continuum field theory approaches, as Schwinger-Dyson equations (SDEs) or Functional Renormalization Group (FRG), and those based on the lattice regularization. In particular, the unveiling of key properties for the nonperturbative behavior of two- and three-point QCD Green's 
functions~\cite{Aguilar:2008xm,Boucaud:2008ky,Fischer:2008uz,
Szczepaniak:2001rg,Szczepaniak:2003ve,Aguilar:2004sw,Boucaud:2005ce,Fischer:2006ub,Kondo:2006ih,
Binosi:2007pi,Boucaud:2007hy,Epple:2007ut,Dudal:2007cw,Dudal:2007cw,Binosi:2008qk,Szczepaniak:2010fe,RodriguezQuintero:2010wy,
Watson:2010cn,Watson:2011kv,Pennington:2011xs,Kondo:2011ab,Boucaud:2011ug,
Bogolubsky:2009dc,Oliveira:2009eh,Cucchieri:2010xr,Ayala:2012pb,Duarte:2016iko,
Cucchieri:2006tf,Cucchieri:2007md,Cucchieri:2008qm,
Tissier:2010ts,Tissier:2011ey,Aguilar:2013vaa,Pelaez:2013cpa,Blum:2014gna,Eichmann:2014xya,Athenodorou:2016oyh,Cyrol:2016tym,Boucaud:2017obn} 
becomes a main pathways' milestone, specially in connection with the emergence of a gluon mass~\cite{Cornwall:1981zr,Bernard:1982my,Donoghue:1983fy,Philipsen:2001ip}, settling thus profound implications~\cite{Binosi:2009qm,Aguilar:2009nf,Aguilar:2010gm,Binosi:2016nme,Gao:2017uox}, 
and leading also to the grounding of a symmetry-preserving truncation of SDEs defining a tractable continuum bound-state problem able to reproduce the observable properties of hadrons\cite{Maris:2003vk,Chang:2009zb,Chang:2011vu,Qin:2011dd,Qin:2011xq,Bashir:2012fs,Eichmann:2012zz,
Roberts:2015lja,Binosi:2014aea}. 
The nonperturbative running of the two-point gauge-field (gluon and ghost) Green's functions has been also exploited in the aim of testing the Operator Product Expansion (OPE) in QCD~\cite{Boucaud:2000nd,Boucaud:2001st,Boucaud:2005xn}, and then identifying fundamental parameters as $\Lambda_{\overline{\rm MS}}$ from lattice QCD computations~\cite{Becirevic:1999uc,Becirevic:1999hj,Sternbeck:2007br,Boucaud:2008gn,Blossier:2010ky,Blossier:2011tf,Blossier:2012ef,Blossier:2013ioa}.

The key importance of these outcomes makes particularly relevant a careful examination of the impact of regularization artifacts on the lattice QCD Green's functions in Landau gauge, as the one performed in Ref.~\cite{Duarte:2016iko}. The role played by the discretization artifacts, crucial for the extrapolation of results to the continuum limit and for the extraction of QCD parameters therefrom, was not fully understood from the analysis therein performed. Two following publications, a ``{\it Comment-to}"~\cite{Boucaud:2017ksi} and its corresponding ``{\it Reply}"~\cite{Duarte:2017wte}, elaborated further on this issue, mainly in connection with the problem of the lattice scale setting, but failed to settle properly the question about continuum extrapolation. 

The object of this paper is, precisely, the thorough examination of the scaling violations for Landau-gauge gluon and ghost propagators, after MOM renormalization, and for the running coupling in Taylor-scheme that can be computed from them. Indeed, an ace of our analysis comes from the deep connection between the bare propagators and the renormalized coupling: the latter results when the former are appropriately combined and multiplied by the bare gauge coupling, which is a parameter of the discretized action fixed by the lattice set-up~\cite{vonSmekal:1997is,Boucaud:2008gn}. Therefore, a consistent analysis of the three quantities is very demanding and we capitalize on the successful description of results obtained from several large-volume lattices, with different set-up's, made only possible by the use of the quenched approximation. We can thus get an insightful understanding of the regularization cut-off effects for gluon and ghost Green's functions, paving the way towards a reliable and very precise extrapolation to the continuum limit.

\section{\label{sec:GreenF} Discretization artifacts on renormalized Green's functions}

Aiming first at arguing on general grounds, let $\Gamma(k^2,a)$ be the bare dressing of the QCD two-point Green's function for either the gluon or the ghost gauge fields; $k$ being the propagated momentum and $a$ standing for a regularization cut-off which drops out by approaching zero ({\it e.g.}, in lattice regularization, $a$ is the length-dimension lattice spacing; while in dimensional regularization, it corresponds to the dimensionless parameter $\varepsilon=(4-d)/2$). 
The renormalized dressing function will be then obtained by applying first a well-defined subtraction procedure, implying a particular prescription, and removing next the cut-off. Namely,   
\begin{equation}
\Gamma_R(k^2,\zeta^2) =  \lim_{a \to 0} \Gamma_L(k^2,\zeta^2;a) = \lim_{a \to 0} Z^{-1}_\Gamma(\zeta^2,a) \Gamma(k^2,a) \ ;
\end{equation}
where $Z_\Gamma(\zeta^2,a)$ is the renormalization constant, defined for a given scheme and at the subtraction point $k^2=\zeta^2$. 
If MOM prescription is considered, any renormalized two-point Green's function is required to take its tree-level value at the subtraction point, {\it i.e.} to amount to unity, such that $Z_\Gamma(\zeta^2,a) \equiv \Gamma(\zeta^2,a)$ and
\begin{equation}\label{eq:GammaL}
\Gamma_L(k^2,\zeta^2;a) = \frac{\Gamma(k^2,a)}{\Gamma(\zeta^2,a)} = \Gamma_R(k^2,\zeta^2) + {\cal O}(a^2) \ .       
\end{equation}
Thus, it becomes manifest that, without taking the explicit limit making the cut-off to drop out, the subtraction procedure cannot generally prevent from the remaining of a residual dependence on the cut-off. Specially in lattice computations, where simulations are carried out in lattices for which the cut-off is fixed by their discretization spacing, one should either try an extrapolation to the continuum limit by the examination of several appropriate simulations, or somehow care about such a residual cut-off-dependence. Let us specialize, for illustrative purposes, to the quenched gluon-propagator dressing function, denoted as $D$, which reads at the perturbative one-loop level and in lattice regularization as follows  
\begin{eqnarray}
D(k^2,a) &=& \left\{ 1 - \gamma_0 \frac{g_0^2(a)}{16 \pi^2} \ln{a^2 k^2} + {\cal O}(g_0^4) \right\} \nonumber \\
&\times& \left( 1 + {\cal O}(a^2 k^2) + H(4) \right) \ ,     
\label{eq:Db}
\end{eqnarray}
where $\gamma_0=13/2$ corresponds to the one-loop anomalous dimension and $g_0$ is the bare gauge coupling, related to the lattice spacing. Apart from the ${\cal O}(a^2 k^2)$-terms, respecting the $O(4)$ symmetry, there also appear other more complicate lattice artifacts, owing to its breaking into the symmetry under the action of the isometry group $H(4)$ of hypercubic lattices on the discrete momentum $a k_\mu \equiv 2 \pi/N n_\mu$, with $n_\mu$ integers and $N$ the lattice volume in units of the lattice spacing, $a$. These higher-order artifacts, indicated explicitly in \refeq{eq:Db}, can be generally expressed in terms of the invariants $a^n k^{[n]} = a^n \sum_{\mu=1}^4 k_\mu^n$ for $n=4,6,8$. In particular, a leading $H(4)$ correction, at the $O(a^2)$-order, is proportional to $a^2 k^{[4]}/k^2$~\cite{Becirevic:1999uc} and is already present in the tree-level gluon propagator computed in a lattice. As will be pointed below, the $O(4)$-breaking artifacts can be efficaciously cured by applying the so-called $H(4)$-extrapolation~\cite{Becirevic:1999uc,Becirevic:1999hj,deSoto:2007ht} and we will then focus here on dealing with the leading artifacts preserving the $O(4)$-symmetry. 

Therefore, after renormalization but before considering the continuum limit, according to \refeq{eq:GammaL} with $\Gamma \equiv D$, one would be left with
\begin{eqnarray}
D_L(k^2,\zeta^2;a) &=& \left\{ 1 - \gamma_0 \frac{g_R^2(\zeta^2)}{16 \pi^2}  \ln\frac{k^2}{\zeta^2} + {\cal O}(g_R^4) \right\} \nonumber \\
&\times& \left( 1 + c_D \; a^2 \left( k^2 - \zeta^2 \right) + o(a^2) \right) \ , \label{eq:DLa} 
\label{eq:1loop}
\end{eqnarray}
where we have also assumed that $H(4)$-extrapolation has been applied and the $O(4)$-invariant corrections have been removed. There, $c_D$ is a constant and $g_R$ is the renormalized gauge coupling. \refeq{eq:DLa} makes explicit a property of the residual Green's functions' cut-off-dependence, implicit from \refeq{eq:GammaL}, namely its vanishing at $k^2=\zeta^2$ for so to enable both conditions simultaneously: $\Gamma_L(k^2,\zeta^2;a) = \Gamma_R(k^2,\zeta^2)=1$. Beyond perturbation theory, a nonperturbative resummation for the expansion in the coupling at a fixed cut-off would lead to a more general expression that, extended by analogy also to the ghost case, would read
\begin{equation}\label{eq:CGamma}
\Gamma_L(k^2,\zeta^2;a) = \Gamma_R(k^2,\zeta^2) \left( 1 + a^2 \; C_\Gamma(k^2,\zeta^2) + o(a^2) \right) \; ;    
\end{equation}
where $\Gamma\equiv D$ (gluon) or $F$ (ghost), and $C_\Gamma$ is a squared-mass dimension function vanishing at $k^2=\zeta^2$. The nonperturbative emergence in the expansion of \refeq{eq:CGamma} of mass scales, as $\Lambda_{\rm QCD}$ or the gluon mass, enables that momentum and lattice spacing decouple and we can thus extend the one-loop result of \refeq{eq:1loop} and conjecture that 
\begin{equation}\label{eq:Clog}
C_\Gamma(k^2,\zeta^2) = c_\Gamma \; (k^2-\zeta^2) + d_\Gamma m_g^2 \; \ln\frac{k^2+m_g^2}{\zeta^2+m_g^2}     
\end{equation}
is an effective description of the leading $a^2$-contribution for the residual cut-off-dependence, where $m_g$ stands for the gluon mass which has been strongly argued to emerge as a result of the so-called Schwinger mechanism to saturate the gluon propagator at vanishing momentum and cure the running coupling from the Landau pole~\cite{Cornwall:1981zr,Bernard:1982my,Donoghue:1983fy,Philipsen:2001ip,Binosi:2009qm}. 

A particular fruitful MOM-like renormalization scheme for the strong running coupling results from the three-point ghost-ghost-gluon Green's function, defined in Landau gauge and at a subtraction point for which the incoming ghost momentum vanishes~\cite{vonSmekal:1997is,Boucaud:2008gn}. Namely, the so called Taylor-scheme running coupling, which has been recently shown to be intimately related to the quark-gap-equation interaction kernel in Dyson-Schwinger approach~\cite{Binosi:2016xxu} and to a process-independent effective charge built in analogy to the Low-Gell-Mann QED charge~\cite{Gao:2017uox}. The particularities of the Taylor-scheme kinematics and the Landau gauge makes the coupling to rely only on the ghost and gluon two-point Green's functions~\cite{Sternbeck:2007br,Boucaud:2008gn}. It specifically reads
\begin{equation}\label{eq:alphaT}
\alpha_T(k^2) =  \lim_{a \to 0} \alpha^L_T(k^2;a) = \lim_{a \to 0} \frac{g^2(a)}{4\pi} F^2(k^2,a) D(k^2,a)      
\end{equation}
wherefrom, applying \refeq{eq:GammaL} and \refeq{eq:CGamma}, the following renormalization flow can be thereupon concluded
\begin{eqnarray}\label{eq:flow}
\frac{\alpha^L_T(k^2;a)}{\alpha^L_T(\zeta^2;a)} &=& F^2_L(k^2,\zeta^2;a) D_L(k^2,\zeta^2;a)  \\
&=& \frac{\alpha_T(k^2)}{\alpha_T(\zeta^2)} \left\{ \rule[0cm]{0cm}{0.5cm} 1 + a^2 \left( 2 C_F(k^2,\zeta^2) + C_D(k^2,\zeta)  
\right) \right.
\nonumber \\ 
&& \rule[0cm]{4.35cm}{0cm} \left. \rule[0cm]{0cm}{0.5cm} + \; o(a^2) \right\} \ .
\nonumber
\end{eqnarray}
Therefore, the conjecture expressed by \refeq{eq:Clog}, translated to \refeq{eq:flow} allows for a simple separation of $k$- and $\zeta$-dependence so that one is left with
\begin{equation}\label{eq:alphaTL}
\frac{\alpha_T^L(k^2;a)}{\alpha_T(k^2)} =  
1 + a^2 \left( c_\alpha k^2 + d_\alpha m_g^2 \ln\frac{k^2+m_g^2}{\Lambda^2} \right) + o(a^2) \ , 
\end{equation}
where $c_\alpha=2 c_F + c_D$ and $d_\alpha = 2 d_F+d_D$, $\Lambda$ being a mass-dimension parameter which can be derived from a nonperturbative subleading $O(a^2)$-contribution in the bare Green's functions that cancels after MOM renormalization in \refeq{eq:GammaL}, thus not  spoiling the condition at $k^2=\zeta^2$, but does not cancel in the running coupling definition, \refeq{eq:alphaT}. 

\section{\label{sec:lat} Lattice data: the analysis}

In what follows, we will examine the validity of \refeq{eq:alphaTL}, and the underlying conjecture about the residual dependence on the cut-off expressed by \refeqs{eq:CGamma}{eq:Clog}. For so to accomplish, we will directly obtain $\Gamma_L(k^2,\mu^2;a)$, and so $\alpha^L_T(k^2;a)$, from several different lattice simulations with different set-up's parameters and evaluate then whether these quantities relate to their continuum counterparts, $\Gamma_R(k^2,\zeta^2)$ and $\alpha_T(k^2)$, as equations suggest. 

\subsection{\label{subsec:set-up} Set-up's and scale setting}

We have produced the $SU(3)$ lattice gauge field configurations $U_\mu(x)$ from the Monte Carlo sampling using the standard Wilson gauge action, 
\begin{equation}
 S_g =  \frac{\beta}{3}\sum_x  \sum_{\substack{
      \mu,\nu=1\\1\leq\mu<\nu}}^4 \left[1-\re\,\tr\,(U^{1\times1}_{x,\mu,\nu})\right] \ ,
  \label{eq:Sg}
\end{equation}
where $\beta\equiv 6/g_0^2(a)$; and next gauge fixed them to the minimal Landau gauge as explained, for instance, in Ref.~\cite{Ayala:2012pb}. The set-up's parameters can be found in Tab.~\ref{tab:setups}. Then, the gauge field is defined as 
 \begin{equation}
 A_\mu(x+ \hat \mu/2) = \frac {U_\mu(x) -
 U_\mu^\dagger(x)}{2 i a g_0} - \frac13\, \tr\,\frac{U_\mu(x) -
 U_\mu^\dagger(x)} {2 i a g_0}, \label{amu}
 \end{equation}
with $\hat \mu$ indicating the unit lattice vector in
the $\mu$ direction. The two-point gluon Green
function is then computed in momentum space through the following
Monte-Carlo average
 \begin{eqnarray}
 \Delta^{ab}_{\mu\nu}(q)=\left\langle A_{\mu}^{a}(q)A_{\nu}^{b}(-q)\right\rangle ,  
\label{greenG}
\end{eqnarray}
with
 \begin{equation}
A_\mu^a(q)=\frac 1 2 \,\tr\,\sum_x
 A_\mu(x+ \hat \mu/2)\exp[i q\cdot (x+ \hat \mu/2)]\lambda^a \ ,
 \label{amufour}
 \end{equation}
where $\lambda^a$ stands for the Gell-Mann matrices and the trace is evaluated in
color space. Then, the gluon dressing function results from taking the appropriate 
trace of the propagator,
\begin{equation}
D(q^2,a)=\frac{1}{24}\sum_{a,\mu} \Delta^{aa}_{\mu\mu}(q) \ .
\label{eq:LatD}
\end{equation}

On the other hand, the Landau gauge ghost propagator result from the
 Monte-Carlo averages of the inverse of the Faddeev-Popov operator, $M$. 
 Namely,  
 \begin{eqnarray}
 F^{ab}(q^2) &=& \frac 1 V \ \left\langle \sum_{x,y}
 \exp[iq\cdot(x-y)] \left( M^{-1} \right)^{ab}_{xy} \right\rangle \nonumber \\
 &=& \delta^{ab}\frac{F(q^2,a)}{q^2} \; .
  \label{eq:LatF}
\end{eqnarray}  
Thus, \refeq{eq:LatD} and \refeq{eq:LatF} define the bare gluon and ghost dressing function which are obtained as the appropriated projection of the lattice propagators. Statistical errors have been derived by applying the Jackknife procedure. More details of the computations can be found in Refs.~\cite{Boucaud:2005gg,Ayala:2012pb}. 

\begin{table}
\begin{tabular}{c|c|c|c|ccc|c|c}
\hline
$\beta$ & 5.6 & 5.7 & 5.8 & 5.9 & & & 6.02 & 6.202 \\
\cline{2-9}
$N$ & 48 & 40 & 48 & 30 & 48 & 64 & 36 & 48 \\
\cline{2-9}
$V^{1/4}$ (fm) & 11.3 & 7.31 & 6.89 & 3.48 & 5.56 & 7.42 & 3.35 & 3.35 \\
\cline{2-9}
confs. & 890 & 580 & 880 & 420 & 400 & 440 & 420 &  420 \\
\hline
\end{tabular}
\caption{\small Set-up's for the simulations herein exploited. Second and third rows respectively correspond to the lattice volumes in lattice and physical units. For the conversion to physical units, we have proceeded as described below.  
\label{tab:setups}
}
\end{table}

Before applying the MOM prescription to get the renormalized dressing functions, as above stated, the $H(4)$-extrapolation is applied to deal with the $O(4)$-breaking artifacts. The dressing functions, being scalar form factors of two-point Green's functions, computed in lattice QCD, are not invariants under $O(4)$ but under $H(4)$ transformations. Therefore, the prescribed recipe implies the average of results obtained for momenta corresponding to the same $H(4)$ orbit (all the lattice four-momenta with the same $k^{[n]}$ invariants) and, next, an extrapolation towards the continuum limit by the subtraction of the $O(4)$-breaking contributions, fitted as smooth functions of $k^{[n]}$ for all orbits sharing the same $k^2$. More details for the $H(4)$-extrapolation procedure can be found in Refs.~\cite{Becirevic:1999uc,Becirevic:1999hj,deSoto:2007ht}. On top of this, for all the set-up's, we will apply an upper cut in lattice momenta: $k a(\beta) \leq \pi/2$.

\begin{table}[h]
\begin{tabular}{c|ccccc}
$a(\beta)/a(\beta_0)$ & $\beta=5.6$ &  $\beta=5.7$ & $\beta=5.9$ & $\beta=6.02$ & $\beta=6.202$  \\
\hline
\cite{Necco:2001xg} & 1.593 & 1.247 & 0.819 & 0.660 & 0.495  \\ 
 Here & 1.646 & 1.273 & 0.811 & 0.648 & 0.486 \\ 
\hline 
& $a_1$ & $a_2$ & $a_3$ & & \\
\cline{2-4}
\cite{Necco:2001xg} & -1.7331 & 0.7849 & -0.4428 & & \\ 
 Here & -1.7934 & 1.0325 & -0.2509 & & \\
\end{tabular}
\caption{\small Ratios of lattice spacings $a(\beta)/a(\beta_0)$ with $\beta_0$=5.8, obtained from Ref.~\cite{Necco:2001xg} and here by applying the scaling of the gluon dressing function as the setting criterion. These ratios have been used to determine the coefficients $a_j$ from \refeq{eq:ratios}. 
\label{tab:ratios}
}
\end{table}

\begin{figure}[thb]
\centerline{\includegraphics[width=0.99\linewidth]{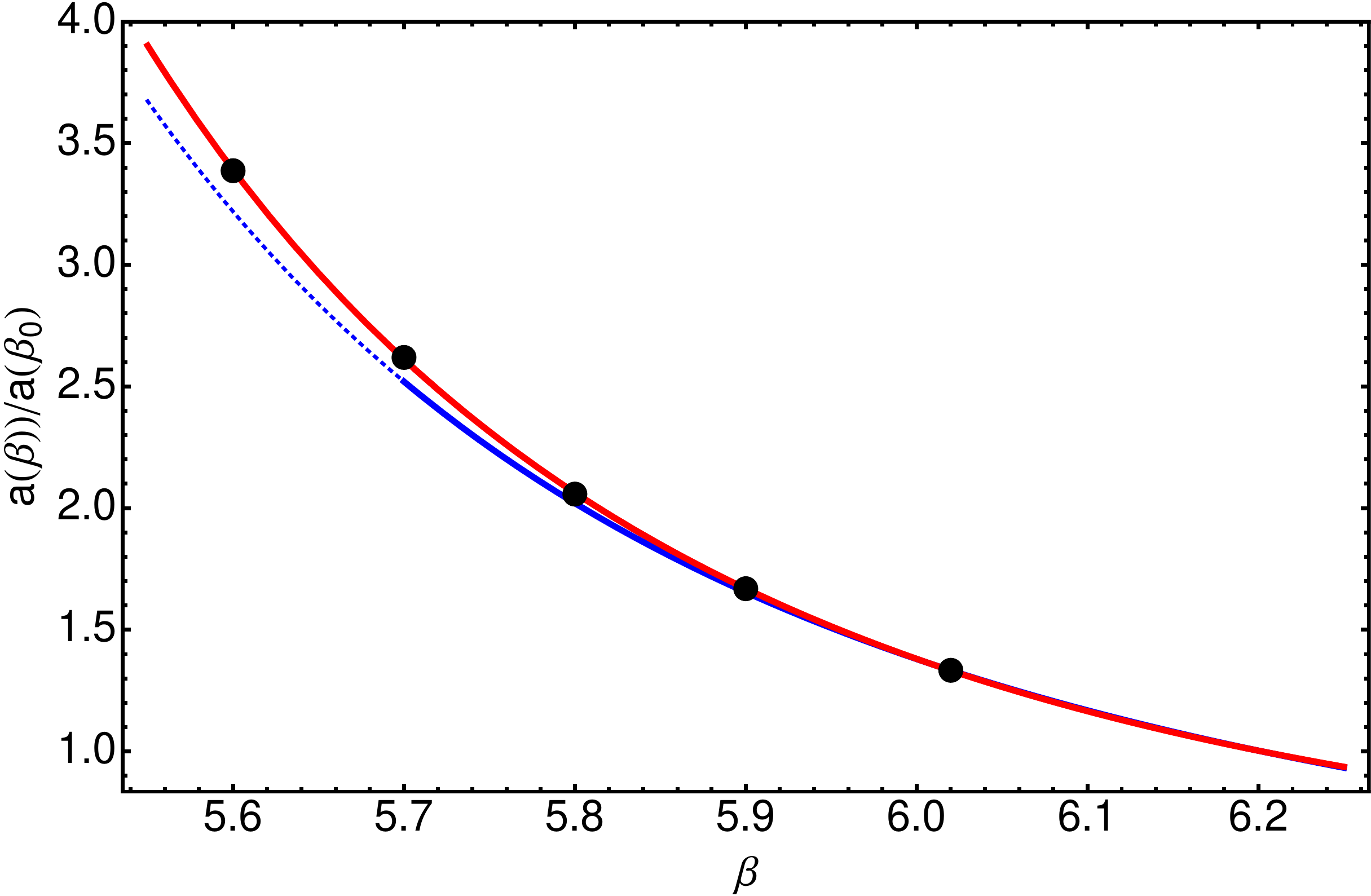}}
\caption{\small  \refeq{eq:ratios} with $a_j$ computed here {\it via} the scaling of the dressing function (red line) and obtained from Ref.~\cite{Necco:2001xg} (blue line). The dotted blue line stands for the extrapolation of results from \cite{Necco:2001xg} outside its original fitting range. In this plot, we have used $\beta_0$=6.202 and applied $a(6.202)/a(5.8)$=0.486, also obtained from the scaling of dressing functions, to convert the ratios of Tab.~\ref{tab:ratios} to the ones herein displayed, $a(\beta)/a(6.202)$ (black solid circles). 
\label{fig:ratios}  
}
\end{figure}

Finally, the scale setting is a key issue for the combined analysis of data resulting from different set-up's with several $\beta$'s. Specially, such a combined analysis relies on an accurate relative {\it lattice calibration}; {\it i.e.}, the knowledge of the ratios of lattice spacings for any pair of $\beta$'s. Indeed, a thorough discussion about how deviations in the lattice calibration might impact on the scaling of renormalized propagators has been the object of a {\it Comment}~\cite{Boucaud:2017ksi} to Ref.~\cite{Duarte:2016iko} and its corresponding {\it Reply}~\cite{Duarte:2017wte}. With this in mind, we can follow Ref.~\cite{Guagnelli:1998ud} and  write
\begin{equation}\label{eq:ratios}
\ln{\frac{a(\beta)}{a(\beta_0)}} = \sum_{j=1}^3 a_j \big\{ (\beta - 6)^j - (\beta_0-6)^j \big\} \; , 
\end{equation}
where the coefficients $a_j$, obtained by a fit of \refeq{eq:ratios} to lattice spacings obtained from applying the Sommer's parameter method for $5.7 \leq \beta \leq 6.92$~\cite{Necco:2001xg}, appear gathered in Tab.~\ref{tab:ratios}. Nevertheless, 
cut-off effects borrowed by the scale-setting procedure, the determination of the heavy quark potential at intermediate distances in the case of the Sommer's parameter, can induce significative deviations at low $\beta$ for the lattice spacings obtained with two different procedures. The effect of these deviations will be anyhow cured by the extrapolation to the continuum limit when is properly made. However, as suggested in Ref.~\cite{Boucaud:2017ksi}, the scaling of a renormalized Green's function obtained for different $\beta$'s, when it exists, provides with a strong criterion guaranteeing the negligible impact on them of cut-effects from the lattice scale setting. This would be an ace for our analysis and we can thus, as done in Ref.~\cite{Boucaud:2017ksi}, assume that $O(4)$-breaking artifacts dominate the cut-off deviations for the gluon propagator so that, after applying $H(4)$-extrapolation, the scaling of the gluon dressing function can be imposed as the condition in order to fix the ratios of lattice spacings. As it will be seen below, this assumption underlying the validity of the relative scale setting, {\it i.e.} the scaling of the gluon dressing function after renormalization and $H(4)$-extrapolation, can be explicitly confirmed {\it a posteriori}. In so doing, we obtain the results of Tab.~\ref{tab:ratios} for the ratios $a(\beta)/a(\beta_0)$ and fit \refeq{eq:ratios} to them, thus obtaining a refined set of coefficients $a_j$, also collected in Tab.~\ref{tab:ratios}, reliable only from $5.6 \leq \beta \leq 6.2$. We have then displayed the results from \refeq{eq:ratios} with the two sets of coefficients $a_j$, from \cite{Necco:2001xg} and the one herein obtained, in Fig.~\ref{fig:ratios} and show that both agree pretty well when $\beta,\beta_0 \geq 5.9$ and that deviations appear only if the ratios involve lower values of the gauge-coupling parameter, distancing a simulation set-up from the continuum limit; the lower is $\beta$ the larger the deviation. Still, if $\beta_0$=6.2, the deviation amounts only a 3.6 \% for ratios with $\beta$=5.7, in the lower border of the validity range of Ref.~\cite{Necco:2001xg}'s results, and increases up to a 5.3 \% with $\beta$=5.6, outside this range. 

In what follows, we will apply the ratios of lattice spacings from Tab.~\ref{tab:ratios} obtained by requiring the scaling of the gluon dressing functions and will thus express all momentum- and mass-dimension quantities in units of $1/a(5.8)$.   A conversion to usual physical units can be done by implementing an absolute calibration for one of the lattices at a particular $\beta$ and applying next the ratios from Tab.~\ref{tab:ratios}. When needed, we will take $r_0\Lambda_{\overline{\rm MS}}$=0.586 and $\ln{(a(\beta)/r_0)}$ from Eq.(2.6) in Ref.~\cite{Necco:2001xg}, $\Lambda_{\overline{\rm MS}}$=0.224 GeV from Ref.~\cite{Boucaud:2008gn} and the ratio $a(\beta)/a(5.8)$ from Tab.\ref{tab:ratios} to obtain $1/a(5.8)=1.372$ GeV and make the conversion to physical units (this particular result is obtained for $\beta=6.2$ but, as it is apparent from Fig.~\ref{fig:ratios}, results derived from any other $\beta$ between 6.0 and 6.2 only differ by less than a 0.5 \%). Now, we are in good position of placing \refeq{eq:alphaTL} for the lattice Taylor coupling under scrutiny. 

\subsection{\label{subsec:aT} The Taylor coupling}

\begin{figure}[thb]
\begin{tabular}{c}
\centerline{\includegraphics[width=0.99\linewidth]{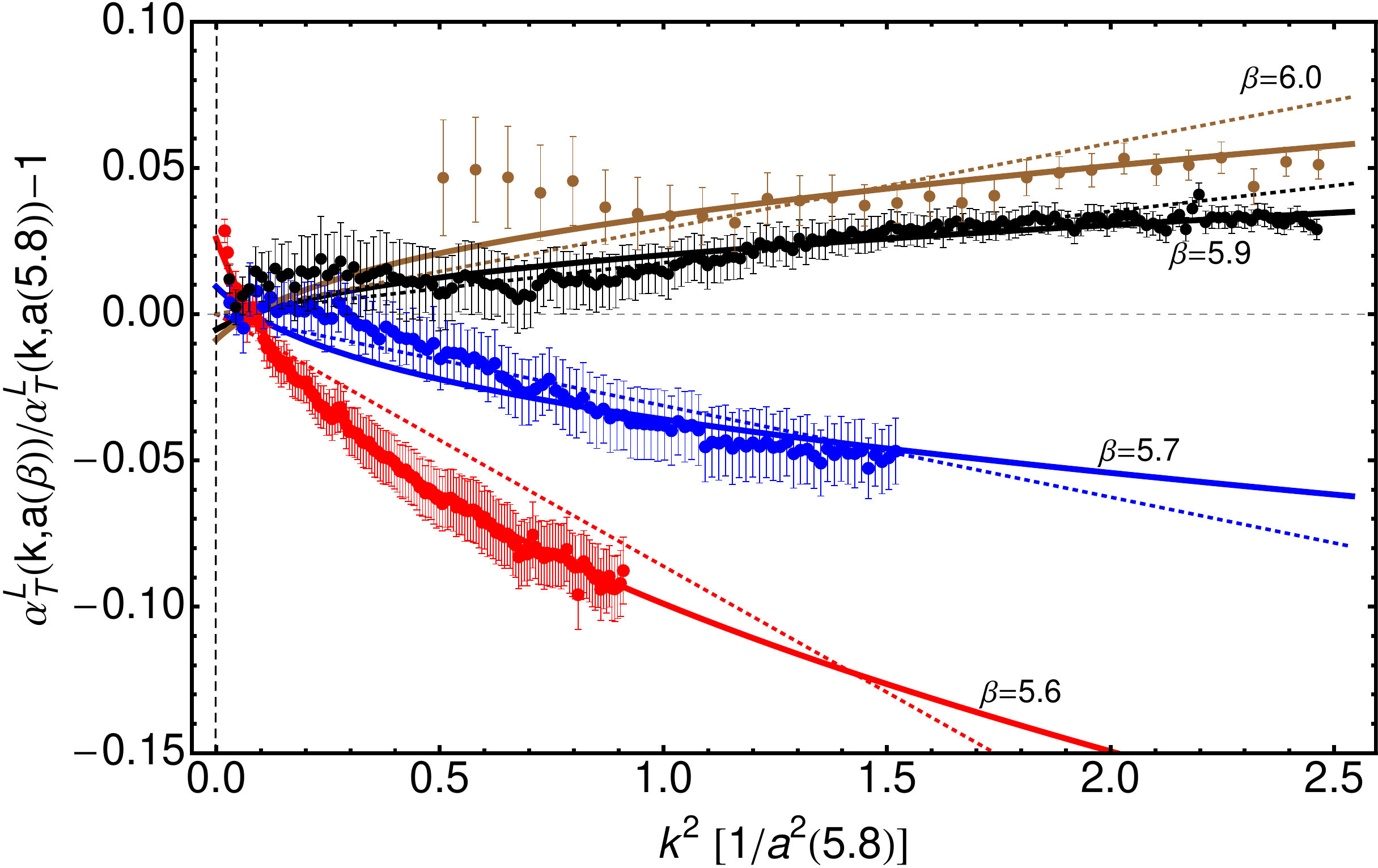}} \\
\centerline{\includegraphics[width=0.99\linewidth]{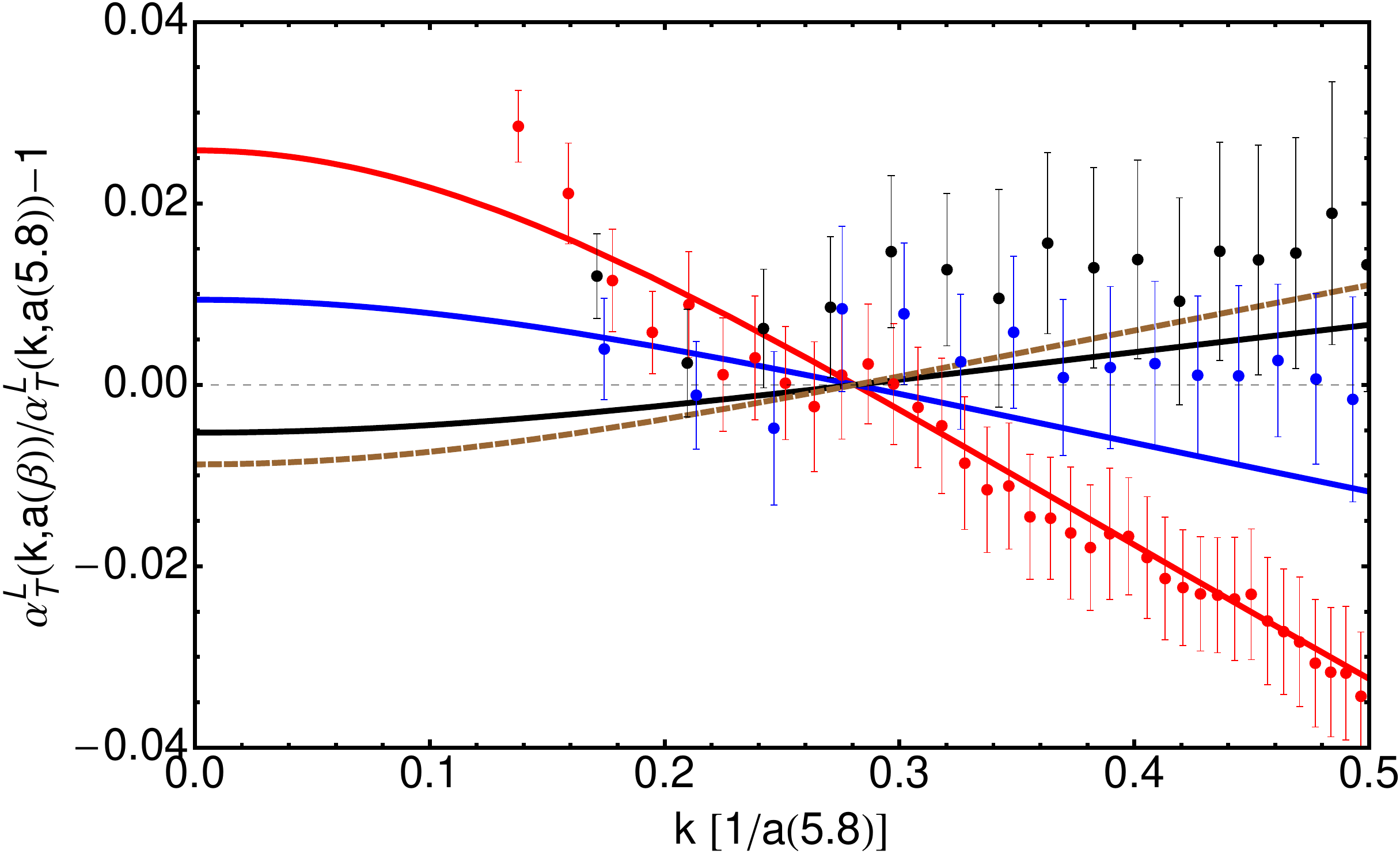}}
\end{tabular}
\caption{\small Lattice data simulated for the set-up's described in Tab.~\ref{tab:setups} and the best-fit to them obtained with \refeq{eq:aTlog} and the parameters collected in Tab.~\ref{tab:comp-aT} (solid lines). In the upper panel, $x$-axis is $k^2$ and the dashed lines correspond to the best-fits with \refeq{eq:aTlog} but with $d_\alpha=0$ (displaying linear corrections in $a^2(\beta) k^2$). All the mass- or momentum-dimension quantities are expressed in units of $1/a(5.8)$. The lower panel is a zoom of the low momentum region aimed at showing the zero-crossing where, for the sake of displaying the feature better, the $x$-axis is chosen to represent $k$ instead of $k^2$.
\label{fig:comp-aT}
}
\end{figure}

In order to need no further assumption concerning the nonperturbative running for the continuum $\alpha_T(k^2)$, we proceed by analyzing ratios of $\alpha_T^L(k^2;a)$ estimated from different lattice set-up's. Indeed, according to \refeq{eq:alphaTL}, these ratios differ from 1 by
\begin{eqnarray}\label{eq:aTlog}
\frac{\alpha_T^L(\overline{k}^2;a(\beta))}{\alpha_T^L(\overline{k}^2;a(\beta_0))} - 1 &=& \left( \frac{a^2(\beta)}{a^2(\beta_0)} -1 \right)  \\ 
&\times& \left( c_\alpha \overline{k}^2 + d_\alpha \overline{m}_g^2\ln\frac{\overline{k}^2+\overline{m}_g^2}{\overline{\Lambda}^2} \right) \ , 
\nonumber
\end{eqnarray}
where the {\it overlined} quantities denote that they are expressed in units of $1/a(\beta_0)$; {\it i.e.}, $\overline{\#} \equiv \# \; a(\beta_0)$. Therefore, the deviations from $1$ for the ratios of $\alpha_T^L(k^2;a)$ appear to be described by an expression, \refeq{eq:aTlog}'s second line, not depending on the lattice parameter's set-up, weighted by the factor $a^2(\beta)/a^2(\beta_0)-1$. Specially, according to our conjecture in \refeq{eq:CGamma}, the parameters $m_g$ or $\Lambda$ should presumably result from a nonperturbative resummmation up to all orders in $g_R$, this is the physical interaction and they will thus be {\it universal} (except for their possible borrowing of higher-order $o(a^2)$-corrections by practical fitting); while $c_\alpha$ and $d_\alpha$ rely on the discretization and will generally depend on the details of the lattice action, the gauge-field definition or the gauge-fixing. However, once these details are fixed and remain the same for all the simulations, as we did, the results obtained for different choices of the bare gauge coupling, $g^2_0(a)=6/\beta$ (see Tab.~\ref{tab:setups}), can only differ by the effect of the factor $a^2(\beta)/a^2(\beta_0)-1$. This is a main feature that is made strikingly apparent by Fig.~\ref{fig:comp-aT}. To produce it, we first compute $\alpha_T^L(k^2;a)$ after \refeq{eq:alphaT} for all the set-up's indicated in Tab.~\ref{tab:setups}.  In particular, we did it for $\beta=5.6,5.7,5.8,6.02$ and the larger volume of $5.9$, and evaluate next \refeq{eq:aTlog}'s l.h.s. with $\beta_0=5.8$, over the momentum intervals where the data for $\beta$ and $\beta_0$ overlap. In order to compute the ratios, the lattice running couplings for $\beta_0$ have been estimated at the same momenta as those for each $\beta$ by performing an interpolation with a Legendre polynomial and propagating errors. Thus, the data displayed in Fig.~\ref{fig:comp-aT} have been obtained directly from the bare gluon and ghost Green's functions without any renormalization other than the multiplication by $6/\beta$, which introduces no additional freedom for data rescaling. As it can be clearly seen, 
\begin{itemize}
\item[(i)] \refeq{eq:aTlog} fits well the data (obtained for five different values of $\beta$), explaining satisfactorily their structure and dispersion in terms of $\beta$, only controlled by the factor $a^2(\beta)/a^2(\beta_0)-1$; 
\item[(ii)] data clearly deviates from a linear behavior on $k^2 a^2(\beta_0)$, strongly supporting the introduction of the \refeq{eq:Clog}'s nonperturbative logarithmic term, and 
\item[(iii)] are consistent with the emergence of a mass scale preventing from the zero-momentum logarithmic divergence. 
\end{itemize}
For the fit, we have taken $m_g$ from Ref.~\cite{Binosi:2016nme} and are so left with the mass-dimension $\Lambda$ and the dimensionless $c_\alpha$ and $d_\alpha$ as free parameters that take the best-fit values gathered in Tab.~\ref{tab:comp-aT}.   
\begin{table}[h]
\begin{tabular}{c|ccc}
$\overline{m}_g$ & $\overline{\Lambda}$ & $-c_\alpha$ & $-d_\alpha$ \\ 
\hline
0.331 & 0.443 & 0.013 & 0.237 
\end{tabular}
\caption{\small Best-fit parameters of \refeq{eq:aTlog} to the lattice data displayed in Fig.~\ref{fig:comp-aT}, obtained for five different ensemble of lattice data with $\beta=5.6,5.7,5.8,5.9$ and $6.02$, with $m_g=0.455$ GeV taken from Ref.~\cite{Binosi:2016nme} and expressed in units of $1/a(5.8)=1.372$ GeV  ($\overline{m}_g$). 
\label{tab:comp-aT}
}
\end{table}

It is worthwhile to highlight a striking feature shown by Fig.~\ref{fig:comp-aT} (specially in the lower panel):  the data for all the different $\beta$ happen to cross zero fairly at the same momentum, thus implying that $\alpha_T^L(\zeta_0^2;a(\beta))=\alpha_T(\zeta_0^2,a(\beta_0))$,  where $\zeta_0$ is the same physical momentum for all $\beta$. In other words, the $O(a^2)$-corrections for $\alpha_T^L(k^2;a)$ in \refeq{eq:alphaTL} become quenched at the same non-zero physical momentum, for which the physical running coupling $\alpha_T(k^2)$ is recovered, irrespective of the values of the gauge coupling and lattice spacing that are used for the lattice set-up. This momentum 
can be estimated as
\begin{equation}
\overline{\zeta}_0 \simeq \left(\overline{\Lambda}^2-\overline{m}_g^2\right)^{1/2} 
\left( 1- \frac{c_\alpha}{2 d_\alpha} \frac{\overline{\Lambda}^2-\overline{m}_g^2}{m_g^2} \right)  \simeq \; 0.29 
\end{equation}
from \refeq{eq:aTlog} and Tab.~\ref{tab:comp-aT}. Actually, Fig.~\ref{fig:comp-aT} tells us that the Taylor coupling directly computed from the bare gluon and ghost Green's functions obtained from four different lattice set-up's ($\beta$ ranging from 5.6 to 5.9) only coincides with each other very much near $\overline{k}=\overline{\zeta}_0$. A fifth simulation at $\beta$=6.02 is performed in a lattice volume so small in physical units that it cannot reach the zero-crossing low-momentum region without being significantly affected by volume artifacts (as will be discussed below). Its data obtained for $\overline{k} > 0.6$ appear however to behave well according to \refeq{eq:aTlog} with the parameters displayed in Tab.~\ref{tab:comp-aT}.  

\subsection{\label{subsec:volume} Continuum limit and finite volume effects}

\begin{figure}[thb]
\includegraphics[width=0.99\linewidth]{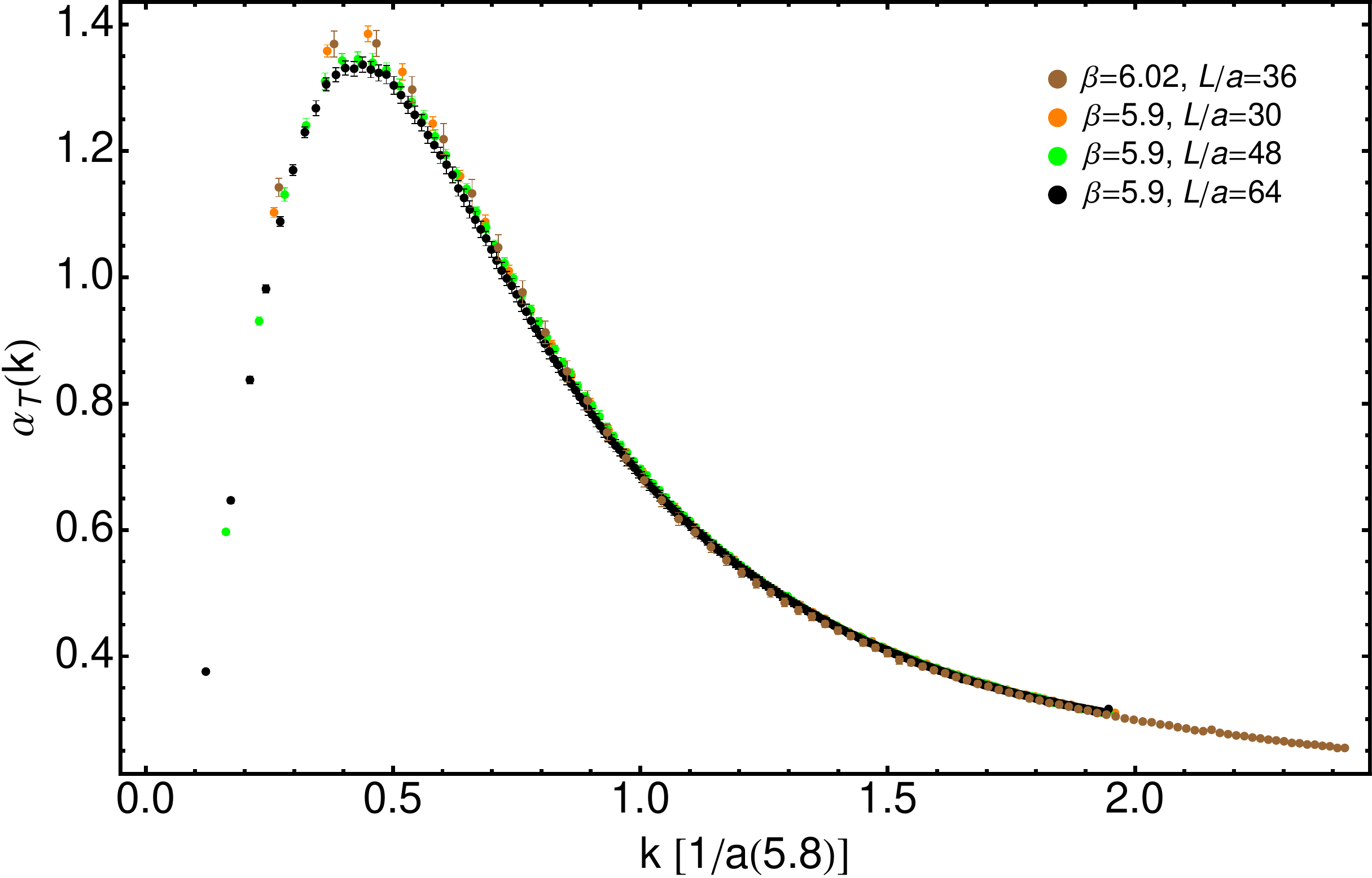}
\caption{\small The continuum Taylor coupling, $\alpha_T(k^2)$, according to \refeq{eq:alphaTL}, with $\alpha_T^L$ computed from four lattice simulations at $\beta$=5.9 and $V^{1/4}=$3.48 ($L/a$=30), 5.56 ($L/a$=48) and 7.42 fm ($L/a$=64) and at $\beta$=6.02 and $V^{1/4}$=3.35 fm ($L/a$=36). 
\label{fig:volume}
}
\end{figure}

After the painstaking scrutiny we have made in the previous subsection for a reliable description of discretization lattice artifacts from the Taylor coupling, \refeq{eq:alphaTL} can be applied to take the continuum limit and thus obtain the nonperturbative physical running of the coupling, $\alpha_T(k^2)$, from its corresponding lattice estimate, $\alpha_T^L(k^2)$. Still, for so to do, one needs to make sure that finite volume effects are under control and do not appear entangled with discretization artifacts in the lattice estimates of the lattice coupling. As it happens in the analysis of an amplitude in spectroscopy, corrections of the order of $\exp{(-mL)}$ are expected for any lattice correlation function, where $L$ is the physical size of the hypercubic lattice and $m$ the mass of the physical bound states which propagates all over the lattice. However, in a quenched theory, even the dominant contribution is negligible when the lattice size is of a few fm's, as it should come from the lightest glueball state, for which $mL \sim O(10)$. On the other hand, when computing the gauge-field correlations functions wich take part in \refeq{eq:alphaT}, a sizeable effect should appear when the associated gauge-field wavelength is, at least, of the same order as the lattice size, {\it i.e.} when $Lp \lwrsim 1$. Thus, the larger is the momentum for which the correlation function is evaluated and the shorter the associated gauge-field wavelength, the less impact the volume effects have. In practice, this impact can be estimated in Fig.~\ref{fig:volume},   where we display the results for the continuum $\alpha_T(k^2)$, obtained with \refeq{eq:alphaTL} and the parameters of Tab.~\ref{tab:comp-aT}, for the lattice estimates from three simulations made at $\beta$=5.9 in three different lattice volumes, $V$=$3.48^4$ ($L/a$=30), $5.56^4$ ($L/a$=48) and $7.42^4$ fm$^4$ ($L/a$=64) and a fourth one at $\beta$=6.02 and $V$=$(3.35\;\mbox{\rm fm})^4$ (see Tab.~\ref{tab:setups}). We can there clearly appreciate that 
\begin{itemize}
\item[(i)] the two simulations made in lattices of $L$=5.56 fm and $L$=7.42 fm at $\beta$=5.9 exhibit results plainly compatible within the errors for all comparable momenta, while the one for the same $\beta$ and $L$=3.48 fm shows a significant volume effect only for $\overline{k} \lwrsim 0.6$;   
\item[(ii)] the results from the two simulations made in the same smaller physical volume and different $\beta$ appear also to fully agree over the whole range of momenta. 
\end{itemize} 
We can thus conclude that the finite-volume artifacts affecting the Taylor coupling {\it via} gauge-field correlation functions are controlled by the lattice physical volume; and that, in quenched QCD, they are negligible in practice when this physical volume is above $(6\;\mbox{\rm fm})^4$. The latter agrees with the findings of Ref.~\cite{Duarte:2016iko}.

\begin{figure}
\begin{tabular}{c}
\includegraphics[width=0.99\linewidth]{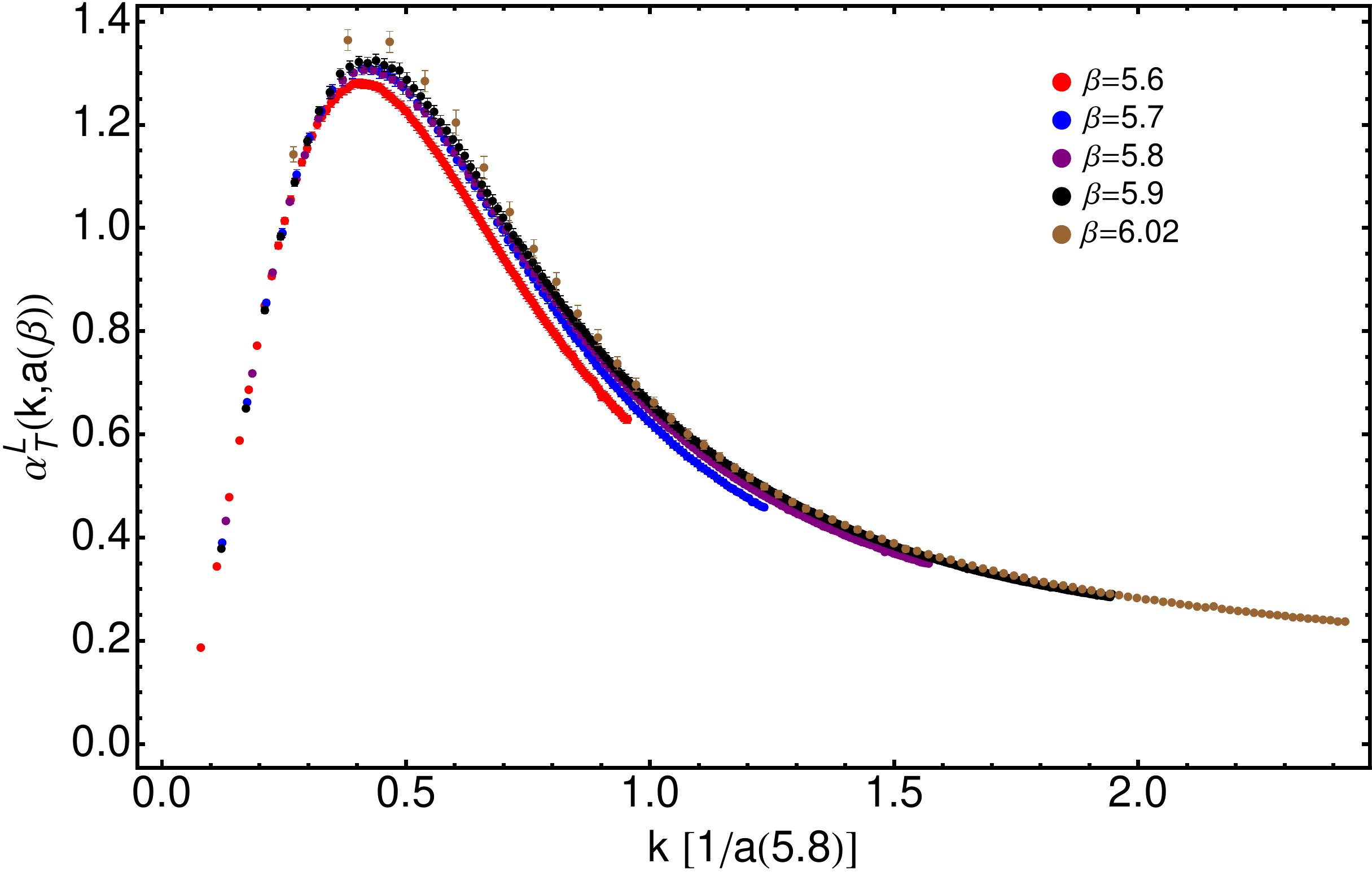} \\
\includegraphics[width=0.99\linewidth]{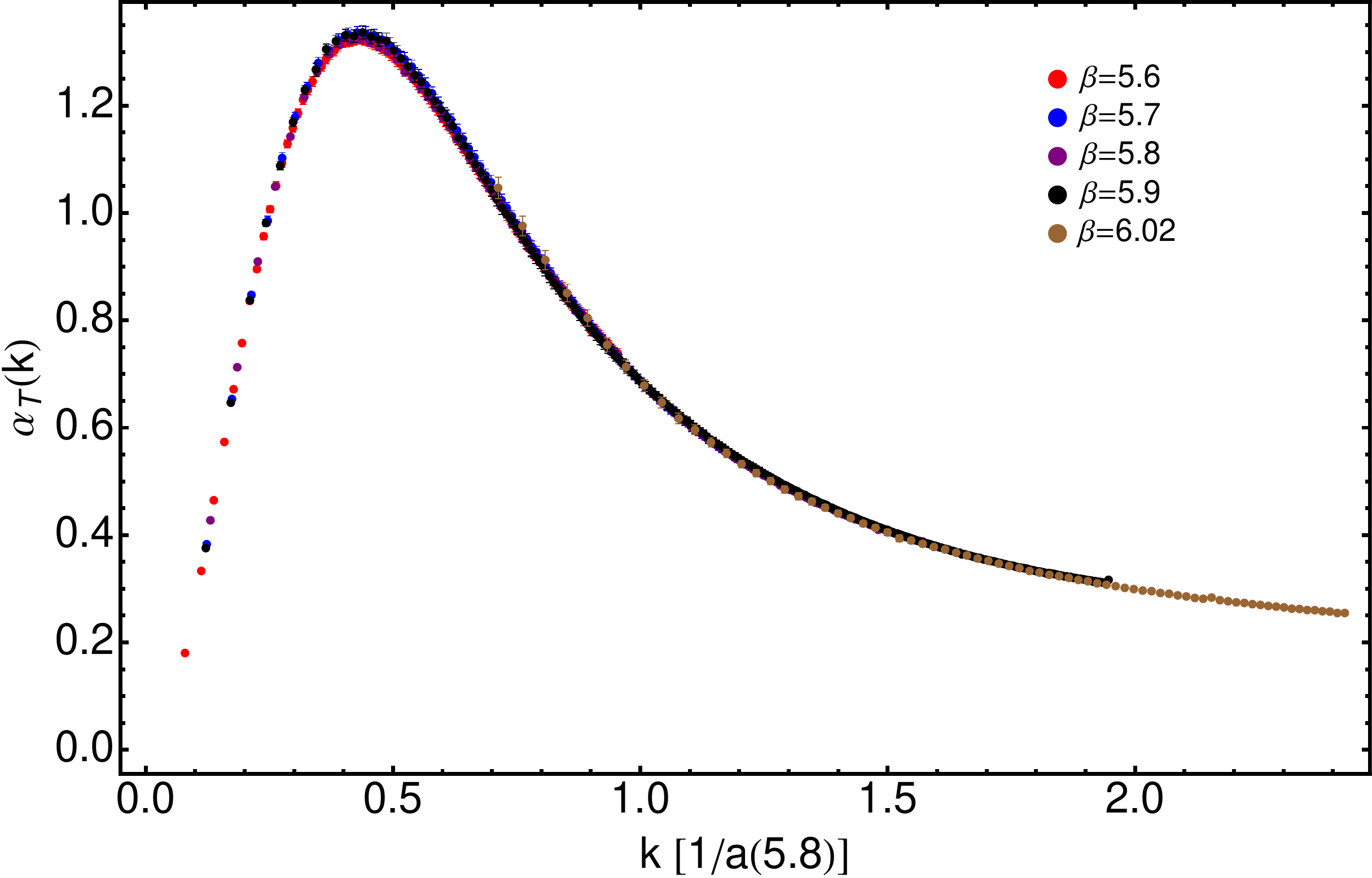} 
\end{tabular}
\caption{\small Lattice estimates for the Taylor running coupling $\alpha_T^L(k^2,a)$ (upper panel) from five different lattice set-up's (see Tab.~\ref{tab:setups}), and their corresponding continuum extrapolations $\alpha_T(k^2)$ according to \refeq{eq:alphaTL} with the best-fit parameters of Tab.~\ref{tab:comp-aT} (lower panel). 
\label{fig:alphaT}
}
\end{figure}

Therefore, the four simulations at $\beta$=5.6,5.7,5.8 and 5.9 we dealt with in Subsec.~\ref{subsec:aT}, made in lattices of more than 6 fm, can be taken, in very good approximation, as free of finite-volume artifacts. In Fig.~\ref{fig:alphaT}, it is shown how the results from these simulations look like, before (upper panel) and after (lower) the continuum extrapolation made through \refeq{eq:alphaTL}. For the sake of comparison, the results obtained at $\beta$=6.02 and in a volume $V$=$(3.35\;\mbox{\rm fm})^4$ lattice appear also displayed and, after extrapolation, all data for $\overline{k} \lwrsim 0.6$ dropped. The scaling found for five different simulations, with five different values of $\beta$, is extremely good, all data lying strikingly on top of each other after applying \refeq{eq:alphaTL} with the gluon mass borrowed from literature and the three other fitted parameters shown in Tab.~\ref{tab:comp-aT}.

\subsection{\label{subsec:props} Gluon and ghost propagators}

\begin{figure}[htb]
\begin{tabular}{r}
\includegraphics[width=0.99\linewidth]{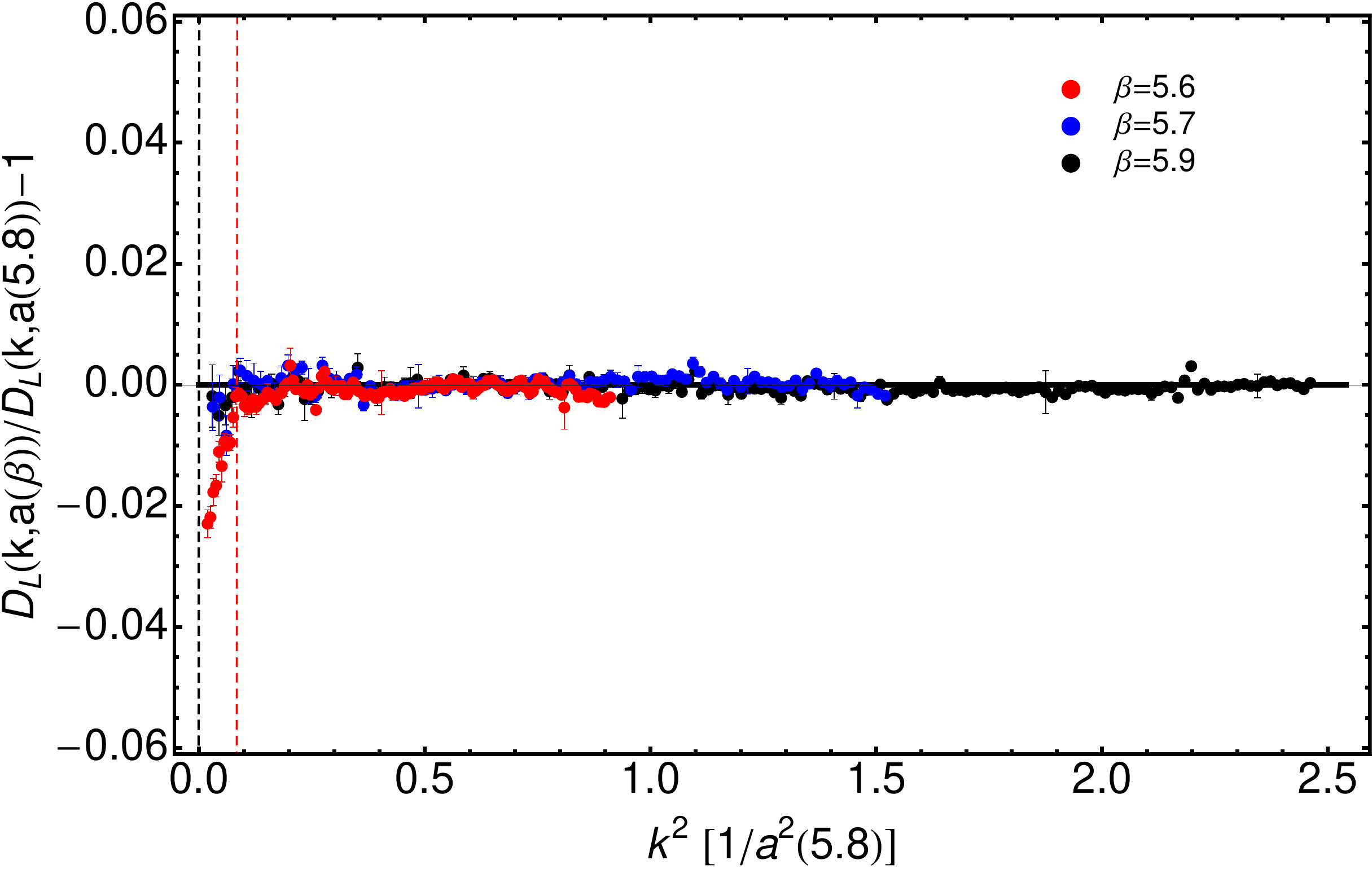} \\
\includegraphics[width=0.925\linewidth]{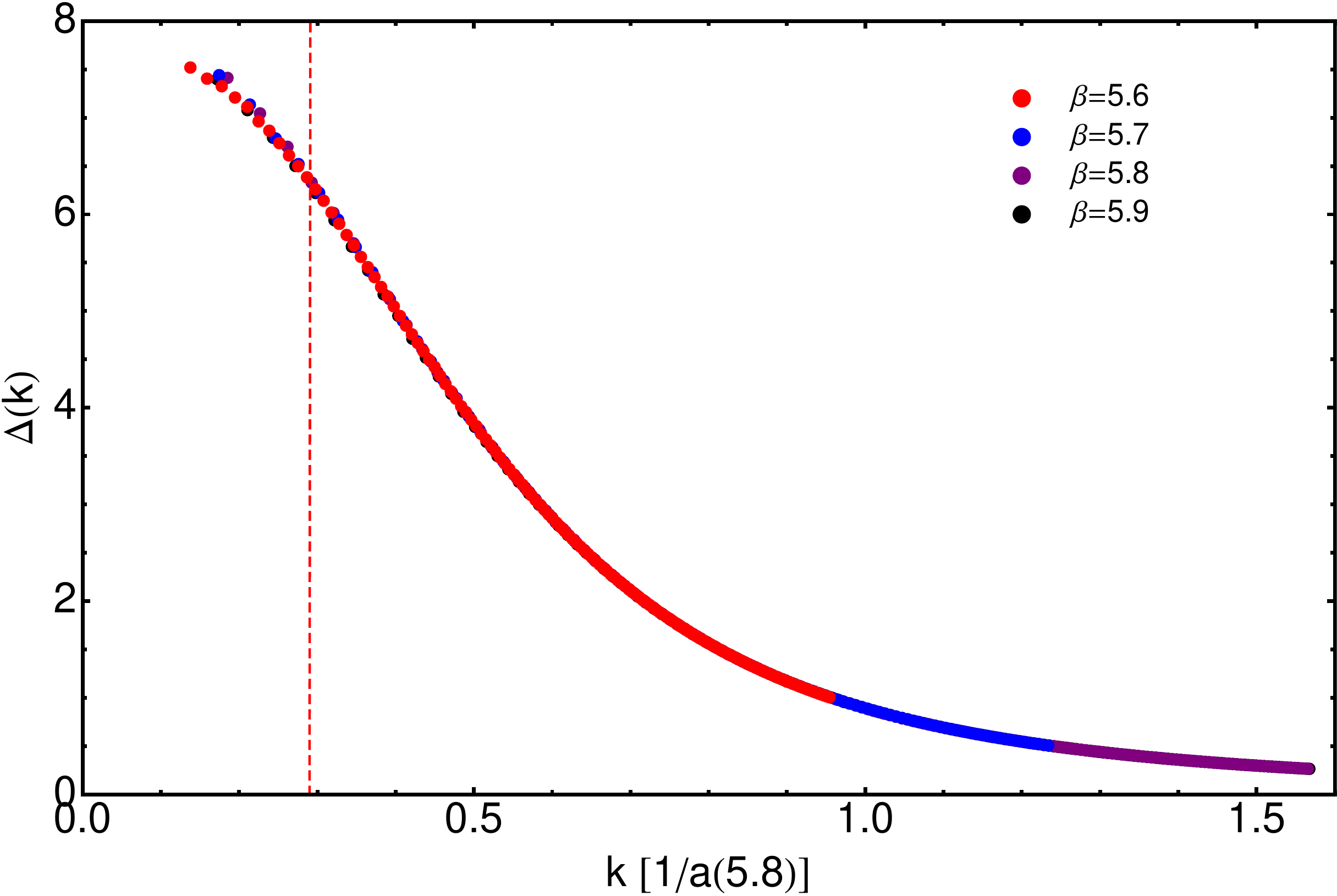} 
\end{tabular}
\caption{\small [Upper panel] The ratios of lattice gluon propagator dressing functions after MOM renormalization, $D_L(k^2;a(\beta))/D_L(k^2;a(\beta_0))$, with $\beta_0$=5.8 and $\beta$=5.6,5.7 and 5.9, are shown to be compatible with $1$ for all momenta except when $k^2 a^2(5.8) \lwrsim$ 0.05, where nearly negligible volume artifacts takes place. [Lower panel] The gluon propagator, $\Delta(k^2)=D_R(k^2)/k^2$, for $\beta$=5.6,5.7,5.8 and 5.9 exhibit a nearly perfect physical scaling. Here, we plotted in terms of $k$ in units of $1/a(5.8)$, instead of the squared momentum, to make more apparent the domain of deeply low momenta. In both cases, the red dashed line is placed at $k a(5.8) \simeq$0.29, to indicate where $\alpha_T^L(k^2;a)$ from all the different simulations crossed. 
\label{fig:Delta}
}
\end{figure}

\begin{figure}[htb]
\begin{tabular}{r}
\includegraphics[width=0.99\linewidth]{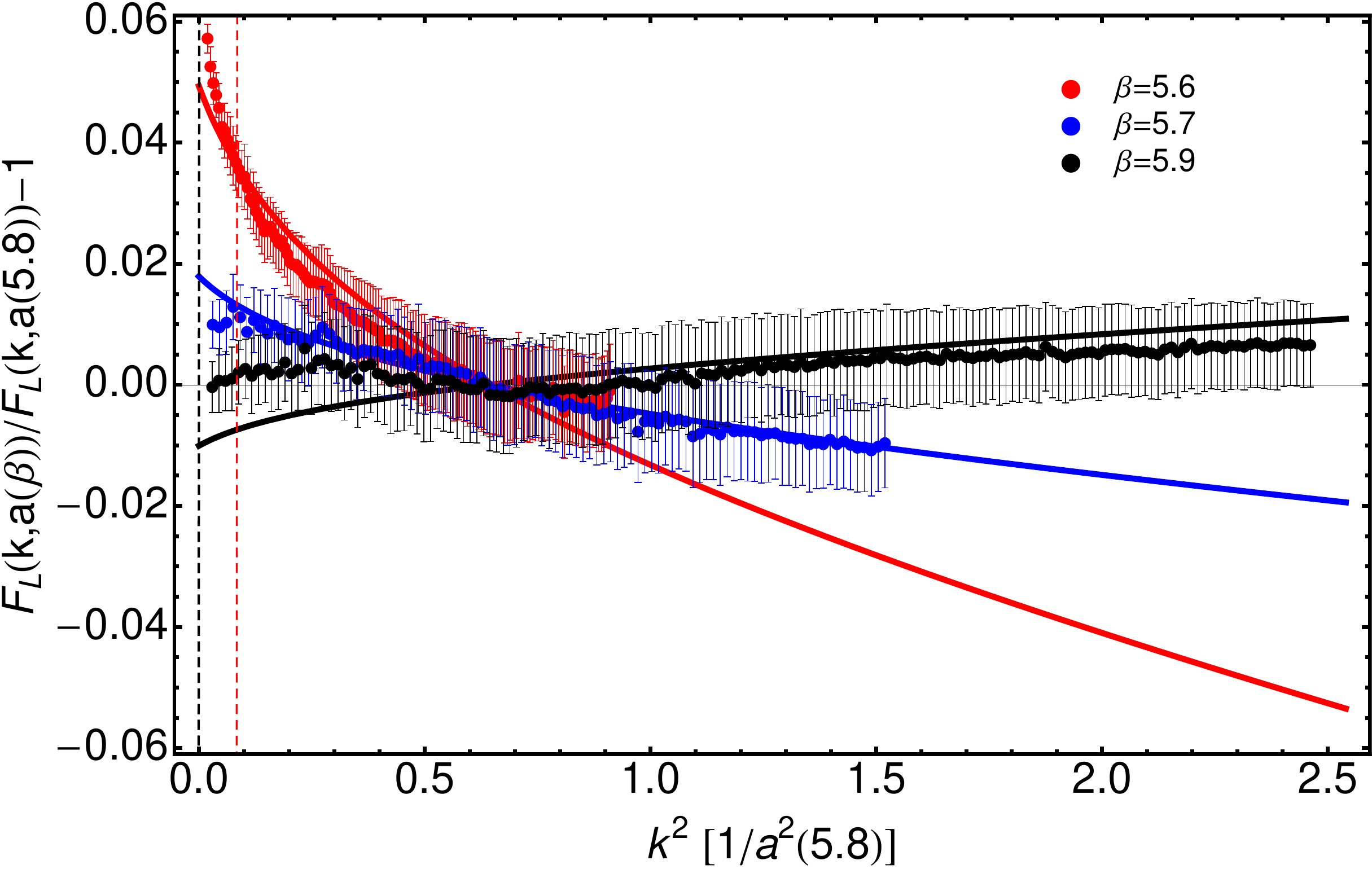} \\
\includegraphics[width=0.94\linewidth]{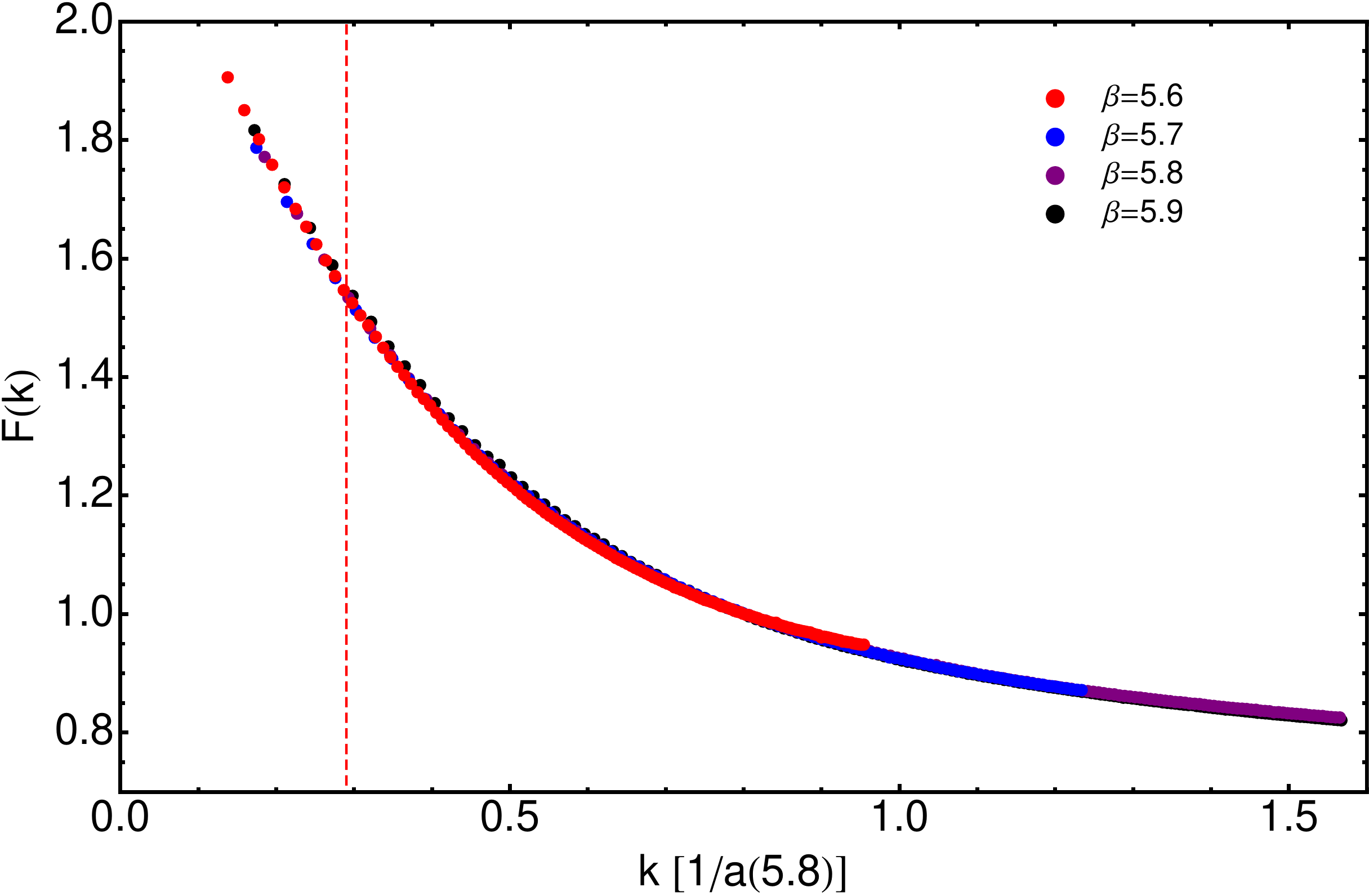} 
\end{tabular}
\caption{\small The ratios of lattice ghost propagator dressing functions after MOM renormalization, $F_L(k^2;a(\beta))/F_L(k^2;a(\beta_0))$, with $\beta_0$=5.8 and $\beta$=5.6,5.7 and 5.9, are shown differ by $1$ according to \refeq{eq:Flog} (solid lines). [Lower panel] After extracting the ghost dressing function $F_R(k^2)$ from simulations at $\beta$=5.6,5.7,5.8 and 5.9 with \refeqs{eq:CGamma}{eq:Clog}, a nearly perfect physical scaling is exhibited. Here, we plotted in terms of $k$ in units of $1/a(5.8)$, instead of the squared momentum, to make more apparent the region of deeply low momenta. In both cases, the red dashed line is placed at $k a(5.8) \simeq $0.29, to indicate where $\alpha_T^L(k^2;a)$ from all the different simulations crossed.   
\label{fig:Ghost}
}
\end{figure}

To obtain the results displayed in the previous subsection for the Taylor running coupling, one essentially needs to deal with bare gauge-field two-point Green's functions and the bare coupling, $g^2(a)=6/\beta$, properly combined as \refeq{eq:alphaT} reads; and the relative lattice calibration described in Subsec.~\ref{subsec:set-up}, which left us with the ratios of lattice spacings of Tab.~\ref{tab:ratios}. We thus apply the $H(4)$-extrapolation to cure from $O(4)$-breaking artifacts and express all the Green's functions, so obtained from different lattice set-up's, in terms of the momentum expressed in the same non-standard but physical units, $1/a(5.8)$; and produce then the plots of Fig.~\ref{fig:comp-aT}. They make strikingly apparent that the remaining cut-off corrections behave as \refeq{eq:alphaTL} dictates. We have employed a relative lattice calibration which agrees well with that in Ref.~\cite{Necco:2001xg} but introduces marginal deviations at low $\beta$ of as much as about 3.6 \%.  This calibration is anyhow grounded on the assumption that $O(4)$-invariant artifacts have a negligible impact on the gluon propagator, implying thereupon that its dressing function should exhibit a physical scaling after renormalization and $H(4)$-extrapolation. 
The latter is a factual statement\footnote{One may argue that the tree-level gluon dressing function in lattice perturbation theory reads $D_L(k^2,a)=1+1/12 \; a^2 k^{[4]}/k^2 + o(a^2)$ and should be so cured only from a $O(4)$-breaking artifact. Furthermore, the examination of the effective operators improving the gauge action {\it \`a la} Symanzik at the ${\cal O}(a^2)$-order seems not to justify the need of curing in addition from $O(4)$-invariant artifacts. However, it is hard to exclude that such artifacts might result in a pure nonperturbative approach.} discussed in Ref.~\cite{Boucaud:2017ksi} and, precisely, applied therein to refine the lattice scale setting. Here, it merely implies that, after MOM renormalization and $H(4)$-extrapolation, both gluon and ghost propagators should behave as dictated by \refeqs{eq:CGamma}{eq:Clog} with
\begin{equation}\label{eq:cs}
c_D=d_D=0 \; , \; \; \: c_F = \frac{c_\alpha} 2 \; , \; \; \; d_F = \frac{d_\alpha} 2 \; ; 
\end{equation}
$c_\alpha$ and $d_\alpha$ given in Tab.~\ref{tab:comp-aT}. This is, strikingly again, confirmed by Figs.~\ref{fig:Delta} and \ref{fig:Ghost}. In them, and owing to a sampling of gauge-field configurations of $O(1000)$, we displayed data for the renormalized gluon and ghost propagators with statistical errors of the order of, respectively, one and five {\it per mil}. Even at this impressive level of statistical accuracy, the ratios of gluon dressing functions obtained in large-volume lattices at four different $\beta$'s do not differ from 1 within the errors, except for deeply low momenta (see upper panel of Fig.~\ref{fig:Delta}). There, for $\overline{k} < 0.2$, results at $\beta$=5.6 and $V^{1/4}$=11.3 fm deviate by around a 2 \% from those at $\beta$=5.7,5.8 and 5.9 and $V^{1/4}\simeq$ 7 fm which, on their side, remain compatible with each other. This strongly suggest that this slight deviation is caused by a still-remaining finite-volume effect. This systematic effect is very nearly negligible, made only apparent by the huge statistics here employed, and only happens at very low momenta: namely, there is no impact from it at $\overline{k} \simeq 0.29$ (highlighted by a red dashed line in Figs.~\ref{fig:Delta} and \ref{fig:Ghost}), the momentum for which all the lattice estimates for the Taylor coupling coincide in Fig.~\ref{fig:comp-aT}. The same is shown in the lower panel of Fig.~\ref{fig:Delta}, where the gluon propagators from the four simulations appear plotted and lie, very accurately, on top of each other. This is a non-obvious result firmly confirming the starting hypothesis of the relative scale setting and the efficiency of the  H4-extrapolation. 

In Fig.~\ref{fig:Ghost}, the ratios of ghost propagator dressing functions from the four lattice simulations are clearly proven to behave according to 
\begin{eqnarray}\label{eq:Flog}
\frac{F_L\left(\overline{k}^2;a(\beta)\right)}{F_L\left(\overline{k}^2;a(\beta_0)\right)} -1 = \left( \frac{a^2(\beta)}{a^2(\beta_0)}-1\right) 
\rule[0cm]{2.5cm}{0cm} \\ 
 \times \; \left( c_F \left( \overline{k}^2 - \overline{\zeta}^2 \right) + c_F \overline{m}_g^2 \ln{\frac{\overline{k}^2+\overline{m}_g^2}{\overline{\zeta}^2+\overline{m}_g^2}}  \right) \; ,
\nonumber 
\end{eqnarray}
obtained from \refeqs{eq:CGamma}{eq:Clog}, and with \refeq{eq:cs} (upper panel); and, after the appropriate continuum extrapolation, the dressing functions are shown to exhibit a nearly perfect physical scaling (lower panel). The renormalization point is chosen here to be $\overline{\zeta}$=0.8, lying thus within the momentum range, and far from its borders, for the four simulations. This is why, precisely, we do not include here the data from the simulation at $\beta$=6.02 in the small volume ($V^{1/4}$= 3.35 fm): they are significantly affected by finite-volume artifacts at this renormalization point and, therefore, the MOM renormalization prescription will contaminate with these artifacts the whole momentum range. Its momentum range and $\beta$=5.8 simulation's offer anyhow a reliable overlap which makes possible the determination of the ratio $a(6.02)/a(5.8)$ in Tab.~\ref{tab:ratios}.

Furthermore, as explained in Fig.~\ref{fig:ratios}'s caption, we have also computed the gluon propagator dressing function for a simulation at $\beta$=6.202 and $V^{1/4}$=3.35 fm, exploited the overlap with $\beta$=5.8 and so extracted the ratio $r=a(6.202)/a(5.8)$ in Tab.~\ref{tab:ratios}. Then, we calculated $a(\beta)/a(6.202)=r^{-1} a(\beta)/a(5.8)$ and plotted the results in Fig.~\ref{fig:ratios}. Indeed, as can be seen in the plot, our estimates for $\beta >$ 5.9 appear to be in nearly perfect agreement with those of Ref.~\cite{Necco:2001xg}.  

It should be reminded that no fit is made here to produce the results displayed in Figs.~\ref{fig:Delta} and \ref{fig:Ghost}. After the scale setting, we merely apply a MOM renormalization prescription, compute the ratios of dressing functions and, when needed, use \refeq{eq:Flog} with the parameters of Tab.~\ref{tab:comp-aT}, obtained for the Taylor coupling,  and \refeq{eq:cs}. It is important to highlight that the connection between the lattice Taylor coupling and the dressing functions relies on the field Theory and its renormalization: \refeq{eq:alphaT} for $\alpha_T^L(k^2;a)$ involves the bare two-point Green functions and the bare gauge coupling which is directly given by the parameter $\beta$ in the lattice gauge action. Their dependence on the lattice spacing is singular for each but cancels in \refeq{eq:alphaT}, remaining only a residual one which vanishes in the continuum limit. Such a residual dependence is also related to the one from the MOM renormalized Green's functions. On the other hand, the way in which the lattice spacing relates to $\beta$ depends on the lattice action and determines the physical scaling of quantities from lattice simulations; namely, the dressing functions in the continuum limit. How all this takes place and match is highly non-trivial. This is what we have exposed here.

\section{\label{sec:conclusions} Summary and conclusions}

We have carefully examined the physical scaling violations of two-points gluon and ghost Green's functions, when they are obtained from a fixed cut-off simulation in lattice regularization, after MOM renormalization and before extrapolation to the continuum limit. Specially, we performed a combined analysis of the gauge-field propagators and the Taylor coupling (obtained from them) seeking a consistent description of results from many lattice simulations, with different lattice spacings ranging widely from 0.07 to 0.24 fm. It should be highlighted that, when the physical scale is properly set, the $H(4)$-extrapolation cures efficaciously the gluon propagator from cut-off deviations up to the order ${\cal O}(a^2)$ and, after renormalization, the results thus obtained show a very striking physical scaling. This is not the case either for the ghost propagator or the Taylor coupling, which are affected by sizeable $O(4)$-breaking artifacts. However, we can accurately deal with these artifacts and get thus an insightful understanding of the impact of discretization cut-off effects on the two-point Green's functions, which makes therefore possible and reliable their very precise continuum extrapolation. This will be of very much help in any future work aiming at extracting QCD parameters from these lattice Green's functions, or just at their comparison when they are obtained from lattice set-up's with different discretization spacings. 

It is furthermore remarkable that the violations of the physical scaling herein scrutinized behave, within our approximation order and after $H(4)$-extrapolation, as the squared lattice spacing times a function of the physical momentum saturated by the gluon mass in the IR limit. The latter appears to suggest that the emergence of the gluon mass become also manifest in the nonperturbative cut-off effects, within a renormalization prescription, before removing them by taking the appropriate limit.    

\section*{Acknowledgements} 
\vspace*{-0.1cm}

We thank the support of Spanish MINECO FPA2017-86380-P research project. The authors are indebted to O. Oliveira and A. Sternbeck for fruitful discussions, mainly in the inception of this work, and to J. Papavassiliou, J. Pawlowski and C.D. Roberts during its completion. SZ acknowledges support by the DFG Collaborative Research Centre SFB 1225 (ISOQUANT). Numerical computations have used resources of CINES, GENCI IDRIS  and of the IN2P3 computing facility in France as well as of L-CSC in Germany.




%

\end{document}